\documentclass[preprint2]{aastex}

\begin{document}


\title{Our Sky now and then -- searches for lost stars and impossible effects as probes of advanced extra-terrestrial civilisations}


\author{Beatriz Villarroel\altaffilmark{1}, I\~nigo Imaz\altaffilmark{1} and Josefine Bergstedt\altaffilmark{1}}
\affil{Department of Physics and Astronomy, Uppsala University}




\altaffiltext{2}{beatriz.villarroel@physics.uu.se}

\begin{abstract}
Searches for extra-terrestrial intelligence (SETI) using large survey data often look for possible signatures of astroengineering. We propose to search for physically impossible effects caused by highly advanced technology, by carrying out a search for disappearing galaxies and Milky Way stars. We select $\sim$ 10 million objects from USNO-B1.0 with low proper motion ($\mu$ $<$ 20 milli arcseconds / year) imaged on the sky in two epochs. We search for objects not found at the expected positions in the Sloan Digital Sky Survey (SDSS) by visually examining images of $\sim$ 290 000 USNO-B1.0 objects with no counterpart in the SDSS. We identify some spurious targets in the USNO-B1.0. We find one candidate of interest for follow-up photometry, although it is very uncertain. If the candidate eventually is found, it defines the probability of observing a disappearing-object event the last decade to less than one in one million in the given samples. 
Nevertheless, since the complete USNO-B1.0 dataset is 100 times larger than any of our samples, we propose an easily accessible citizen science project in search of USNO-B1.0 objects which have disappeared from the SDSS.
\end{abstract}

\keywords{astrobiology --- extraterrestrial intelligence --- surveys}

\section{Introduction}\label{sec:intro}

The possibility of finding intelligent life beyond our own planet is an ongoing dream. So far we have not seen 
or heard the slighest hint. Efforts to search for indirect signatures of astroengineering from extra-terrestrial civilisations in survey data are currently expanding and these searches often make use of large data sets. Such studies have the advantage that they can yield new insights into important astrophysical phenomena as an exciting by-product. Some of these efforts have targeted stars in the Milky Way 
\cite[e.g.][]{Jugaku,Timofeev}, while others have turned to extragalactic scales \cite[e.g.][]{Annis,Wright2014a,Wright2014b,Griffith}. Assuming that all advanced Kardashev II-IV civilisations \citep{Kardashev} 
attempt to build Dyson spheres \citep{Dyson} throughout their galaxy to harvest energy from stars, a significant fraction of the waste heat from the Dyson spheres will be irradiated as mid-infrared (MIR) emission.
When looking for high-consuming alien civilisations producing a waste-heat luminosity of $\sim$ 10$^{11}$ $L_{\sun}$ 
from each galaxy, it was shown with Wide-field Infrared Survey (WISE) data \citep{OtherWright} that such super 
civilisations must be extremely rare \citep{Wright2014b}. Recent studies using other methods \citep{Olson,Zackrisson,Lacki} 
support the conclusion that Kardashev II-IV civilisations in the Local Universe are very rare and difficult to find. 
The difficulty to distinguish between mundane causes and the signatures of astroengineering, further complicate 
the searches.

In this paper, we propose replacing the search for possible signatures of astroengineering
with a search for impossible (or nearly impossible) effects for conventional astrophysics.
Quoting Arthur C. Clarke's Third Law, ``Any sufficiently advanced technology is indistinguishable from magic.'' 
Examples of such hypothetical effects include: a galaxy that rapidly and strongly changes redshift or apparent size 
over the course of a few years. Or a galaxy, previously visible on the sky, that suddenly disappears $entirely$ from its 
location. Also a Milky Way star that disappears entirely without an accompanying supernovae is equally interesting and 
can indicate the existence of advanced extraterrestrial civilisation with an interest in 
hiding a star from their enemy. But it can also point towards unknown or exotic physics,
like stars disappearing into wormholes.

We perform a search for objects lost from our Sky over the last decades. A similar idea of 
constructing an anti-transient survey of stars in nearby galaxies to search for the 
hypothesized failed supernovae from massive stars, was originally proposed 
by Kochanek et al. (2008), 
later also carried out \cite[]{FailedSupernovae}. Failed supernovae can only happen to stars within the mass range of 18-25 solar masses and are significantly less common than optically bright supernovae. In 2015, two candidates for failed supernovae 
were detected \citep{Reynolds} in nearby galaxies. It should be noted that not a single supernova in the Milky 
Way has been detected in the last 150 years, therefore the number of failed supernovae should be zero.

In this paper, we conduct a study and present results from a search for advanced extraterrestrial civilisations
by looking for objects that mysteriously disappeared from the sky in the last decade. We use the
United States Naval Observatory (USNO) B1.0 Catalogue catalogue \citep{Monet} that stores 
information about roughly 1 billion objects and is complete down to $v \sim$ 21. The USNO-B1.0 
objects have negligible proper motion ($<$ 20 m.a.s./year) and were all detected during at least two epochs in the 
Palomar Observatory Sky Survey (first POSS survey: 1950-1966; second POSS survey: 1977-1999) before being 
included in the catalogue. This minimizes detections of asteroids, comets, fast-moving objects
and chance detections. Afterwards, we match these objects against a third epoch using the Sloan Digital Sky Survey (SDSS)
 \citep{York2000} and search for objects that have disappeared. The SDSS is complete down to magnitude $r \sim 22$ (ugriz) and overlaps partially in sky coverage with USNO-B. By comparing the USNO-B1.0 and SDSS catalogues, it is shown that USNO-B1.0 is 100\% complete for unblended stars \citep{Monet}.

Any candidate found must be explored with follow-up photometry to separate it from natural anti-transient events, e.g. variable 
objects. (We only consider a object really ``disappeared'' if it does not reappear during
any follow-up observations.) Finally, we use this study to set a limit on the probability for observing a disappearing object-event 
with the wide-field surveys. 

\section{Samples \& Methods}

We use the NASA/IPAC Infrared Science Archive interface to access USNO-B1.0 data\footnote{http://irsa.ipac.caltech.edu}. 
It is not possible to download samples larger than about 10 million objects due to technical limitations on both the server 
page and our own technical capacities. Therefore we construct parent catalogues based on various selection 
criteria. We mark the constraints for each parent sample in Table \ref{Constraints}. One common constraint in
all samples is that we only use objects with low proper motion ($<$ 20 m.a.s. / year).

Variability near the detection limit of the USNO-survey can cause some objects to drop out of the catalogue.
Cataclysmic variables are example of stars that might appear or disappear between the USNO-B and SDSS surveys. But also,
static objects could disappear if they are weak and the background sky is bright. The 1$\sigma$ error in
the r-band magnitude error is about $|\Delta r | \sim $ 0.25 mag, and we define
``static'' objects as those showing no signs of variability between the first and second epochs 
in USNO-B1.0 so that $|\Delta r |< $ 0.25 mag. ``Variable'' objects have $|\Delta r |> $ 0.25 mag.
The NDET and Big samples contain only static objects, NDETVar only variable ones, while the Weak sample contains
both static and variable ones.

\begin{table*}[ht]
\caption{Selection constraints for the parent samples from USNO-B1.0.}
\centering
{\tiny
\begin{tabular}{c c c c c}
\hline\hline
\multicolumn{5}{c}{Samples} \\
\hline\hline
Parameter & NDET (N) & Weak (N) & Big (N) & NDETVar\\
Character? & Static obj. & Any & Static & Variable\\
Ra$_{Lower}$ J2000 & 110 & 110 & 110 & 110\\
Ra$_{Upper}$ J2000 & 265 & 260 & 260 & 265\\
Dec$_{Lower}$ J2000 & -3 & 0 & 0 & -3\\
Dec$_{Upper}$ J2000 & 75 & 70 & 70 & 75\\
Mean epoch & - & year 1949 - 1966 & - & -\\
Number of detections & $>$ 4 & - & - & $>$ 4\\
1$\sigma$-error in right ascension ra$_{err}$ at 1st epoch [100mas] & - & - & $<$ 20 & -\\
1$\sigma$-error in declination dec$_{err}$ at 1st epoch [100mas] & - & - & $<$ 20 & -\\
Proper motion pm$_{ra}$   [mas] & $<$ 20 & $<$ 20 & $<$ 20 & $<$ 20\\
Proper motion pm$_{dec}$   [mas] & $<$ 20 & $<$ 20 & $<$ 20 & $<$ 20\\
Error in proper motion pm$_{ra,err}$   [mas] & - & $<$ 1 & $<$ 10 & -\\
Error in proper motion pm$_{dec,err}$   [mas] & - & $<$ 1 & $<$ 10 & -\\
1st red observation r$_{epoch-1}$ [mag] & $<$ 18 & 11 $< r < $ 20 & $<$ 17.5 & $<$ 18\\
2nd red observation r$_{epoch-2}$ [mag] & $<$ 18 & 11 $< r < $ 20 & $<$ 18.5 & $<$ 18\\
1st red observation br$_{epoch-1}$ [mag] & $<$ 18 & - & - & $<$ 18 \\
2nd red observation b$_{epoch-2}$ [mag] & $<$ 18 & - & - & $<$ 18\\
Distance (X) between photo-center and mean position 1st epoch $\xi_{r ,epoch-1}$ [arcsec] & - & $<$ 1 & - & -\\
Distance (Y) between photo-center and mean position 1st epoch $\eta_{r ,epoch-1}$ [arcsec] & - & $<$ 1 & - & -\\
Distance (X) between photo-center and mean position 2nd epoch $\xi_{r ,epoch-2}$ [arcsec] & - & $<$ 1 & - & -\\ 
Distance (Y) between photo-center and mean position 2nd epoch $\eta_{r ,epoch-2}$ [arcsec] & - & $<$ 1 & - & -\\[0.2ex]
\hline
\hline
\end{tabular}}
\label{Constraints}
\end{table*}

\begin{table*}[ht]
\caption{Sample sizes after every step of the cleaning. Objects for which no nearest neighbour can be found, or where the nearest neighbour is at least $\omega$ $>$ 0.15 arcmin away, are counted in N$_{Nomatch}$.}
\centering
\begin{tabular}{c c c c c}
\hline\hline
\multicolumn{5}{c}{Samples} \\
\hline\hline
Sample sizes & NDET (N) & Weak (N) & Big (N) & NDETVar(N)\\
N$_{start}$ & 7352942 & 1607582 & 10310181 & 1041330\\
N$_{Footprint}$ & 4921685 & 1305447 & 4906527 & 802295\\
N$_{Nomatch}$ & 186262 & 30566 & 53569 & 17761\\
N$_{Preliminary}$ & 13 & 110 & 25 & 0\\[0.2ex]
\hline
\hline
\end{tabular}
\label{SampleSizes}
\end{table*}

We use the Footprint function in the SDSS Casjobs to cross-match the objects to Sloan Digital Sky Survey Data Release 12 (DR12). 
The Footprint function checks if a coordinate is within the SDSS scanned field. For objects within the scanned field we 
search for the nearest primary neighbour in SDSS DR12 and the angular distance $\omega$ (arcmin) to each object. For 
those objects for which no near primary photometric neighbour can be found in the vicinity, the distance is replaced by ``NULL''. 
We only keep objects where the distance is ``NULL'' or objects that have neighbour at an angular distance $\omega$ $>$ 0.15 arcmin. The sample sizes at the different steps can be found in Table \ref{SampleSizes}.

Using the SDSS Object Explorer, we visually examine one by one in the SDSS Object Explorer (list of targets). 
For the majority of the objects an object is found in the center of the window despite being marked as ``null'', showing they were missed by the SDSS targeting pipeline. We ignore these. We also ignore all showing a black ``dead'' stripe, near blended stars, diffraction spikes
and other visible artifacts in the SDSS. A tiny fraction of the images appear normal without any recognizable artifacts and yet have no object
in the center of our image. These constitute our preliminary objects of interest that might harbour important candidates.

\section{Results}

From the three sample selections we obtain in total 148 preliminary objects (see Table \ref{SampleSizes}) of interest that are missing in the SDSS and have no 
primary photometric object detected in the vicinity of the SDSS images. The number of discovered preliminary objects is also marked in
Table \ref{SampleSizes}.

Most of objects from the Weak sample classification have zero proper motion and weak $r$-magnitude $ r\sim$ 19, close to our selection limit. 
One may wonder whether we are dealing with detection faults and enhanced noise. In order to find out if the reported event truly is an object
lost in the third epoch, we must be sure of two things: first that the objects were truly there from the beginning (removing false positives), 
and second, that they truly were lost (removing false negatives).

\subsection{False positives in USNO-B1.0? Artifacts?}

Artifacts and false detections in USNO-B1.0 seem like the easiest ways to explain
the apparent missing objects. Whatever the artifact is, it must be the type of artifact 
that appears in both the first and the second survey and could appear over and over
again with the same instrumentation. Diffraction spikes and halos near bright stars 
could be present and cause repeated misidentification by the survey machinery.
Scratches and damages on the photometric plug plates, cosmic rays or the spontaneous fly-over
by a swarm of ducks during the photon collection are events with a negligible 
probability of happening during two different epochs for the same spot on the sky.
Luckily, many of the spurious objects in the USNO-B1.0 were identified by a clever
computer algorithm and are stored in a list \citep{ArtifactPaper}. Only a handful of our 
interesting targets are found within 5 arcseconds from listed positions of the spurious targets.

But false positives can also happen if the sky substraction or signal is quite weak
relative to the background noise. Coincidences due to erroneous combinations of objects
with high proper motion at different epochs can also create a false positive.

Therefore, we visually examine the original POSS-1 and POSS-2 images
that were used to construct the first and second epoch of the USNO-B1.0 survey, 
obtainable from the Digitalized Sky Survey. Using the central coordinates of the targets, 
we insert them into the target search and see if any object is visible in the center
of the POSS-1 images. For the majority of the targets we cannot see anything even in POSS-1, suggesting these objects have errors in their proper motions and predicted J2000 positions
or were just noise (and no real detections). But some preliminary candidates survive this check.

\subsection{False negatives in SDSS?}

An object can be lost from the SDSS if the photometry is of lower depth. However,
SDSS goes deeper and many more faint stars are visible in the SDSS Object Explorer than in the POSS images. 

For every object found in the POSS-1 and POSS-2, we must compare the relative position in its field
to the relative position in a SDSS image. We note that almost every object is found in SDSS where it was supposed to be 
from the beginning -- at its expected position, but with disagreement in SDSS and USNO-B1.0
coordinates.

Summarizing the three typical USNO-B1.0-specific errors that seem to be present upon visual inspection in Table \ref{SampleSizes}, we find:

\begin{enumerate}

\item (False positive.) We see no object at all in the center of the original POSS-1 image
or some very weak signal that could be noise. More than half of the errors belong to this category. 
To know the probability of false positives of this nature to occur, it would be advantageous 
to obtain an estimate of how probable it is to get $n$ independent false detections at the
same spot in the sky in both POSS-1 and POSS-2. (This is beyond the scope of this study but 
will be implemented once we redo it with the full USNO-B1.0 catalogue.)

\item (False positive.) We see an object clearly in one survey, but in the other survey
nothing or an artifact (e.g. a stripe) is seen.

\item (False negative.) There are objects in the center of both POSS-1 and POSS-2, but
the coordinates are slightly offset between POSS and SDSS. The number of 
``preliminary candidates'' belonging to this category is about $\sim$ 10\% of the cases.

\end{enumerate}

Not unexpectedly, the number of ``preliminary candidates'' identified drops dramatically
if requiring an object to have at least four detections. The requirement eliminates 
false positives due to faintness and might improve the quality when working with data
from noisy photographic plates. The preliminary candidates originating from the NDET sample almost all have slightly offset coordinates between SDSS and POSS. This demonstrates the need to use samples with at least four detections in the USNO-B1.0 when searching for missing objects. We know that 40\% of all USNO-B1.0 objects at the northern hemisphere have five detections but unfortunately we cannot access the full sample.

\subsection{A dubious candidate}

From the entire study, only one candidate survives our rigorous investigation
and cannot be rejected. It originates from the Weak Sample and has the coordinates 
(ra,dec)=224.402387,18.417250 (J2000) or 14h 57m 36.57s +18d 25m 02.10s. 

\begin{figure*}
 \centering
 \includegraphics[width=0.45\textwidth]{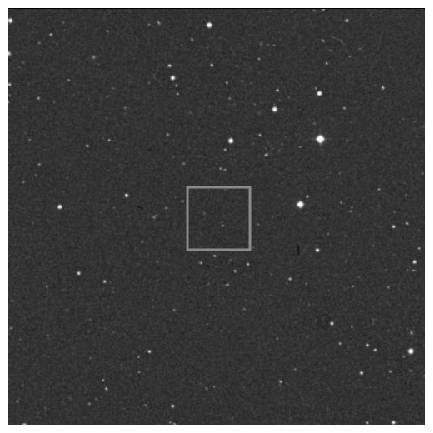}\
 \includegraphics[width=0.45\textwidth]{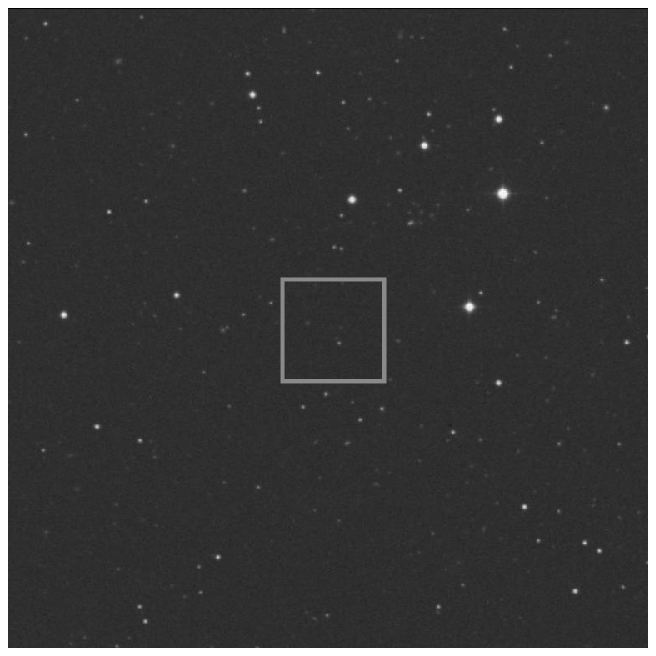}\
 \includegraphics[width=0.45\textwidth]{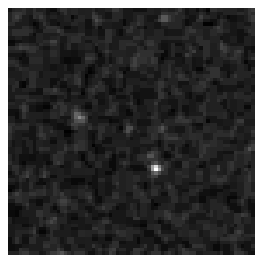}\
 \includegraphics[width=0.45\textwidth]{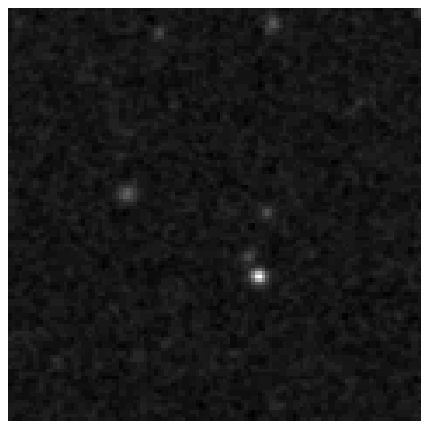}\
 \caption{POSS images of the candidate at two different epochs. The two upper images show images in the POSS1 E filter (1950-03-16, resolution 1.6''/pixel) and POSS2 F filter (1992-03-10, resolution 1.0''/pixel) and are taken from the CDS Portal. The two lower images are zoom-ins of the same objects at two epochs, watched via the STScI server (POSS I Red, POSS II/UKSTU Red). In the first epoch the object is clearly seen. In the second epoch, there are still hints of an object. Note that the two upper images come from different sources than the two lower. We recommend the reader to look at the original images in the fits format available at STScI server or CDS Portal.}
 \label{POSSimages}
\end{figure*}

\begin{figure*}
 \centering
 \includegraphics[width=0.45\textwidth]{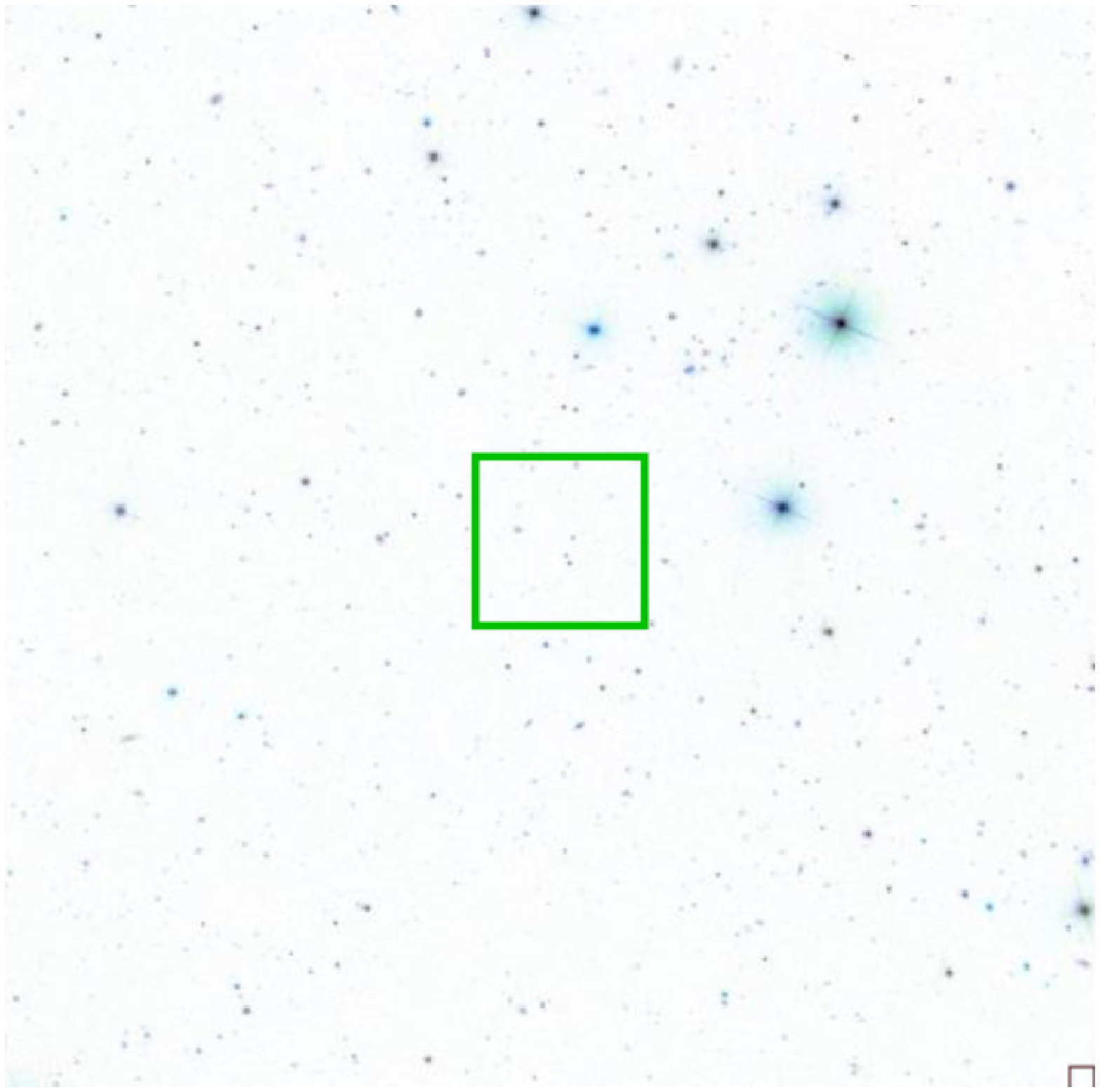}\
 \includegraphics[width=0.45\textwidth]{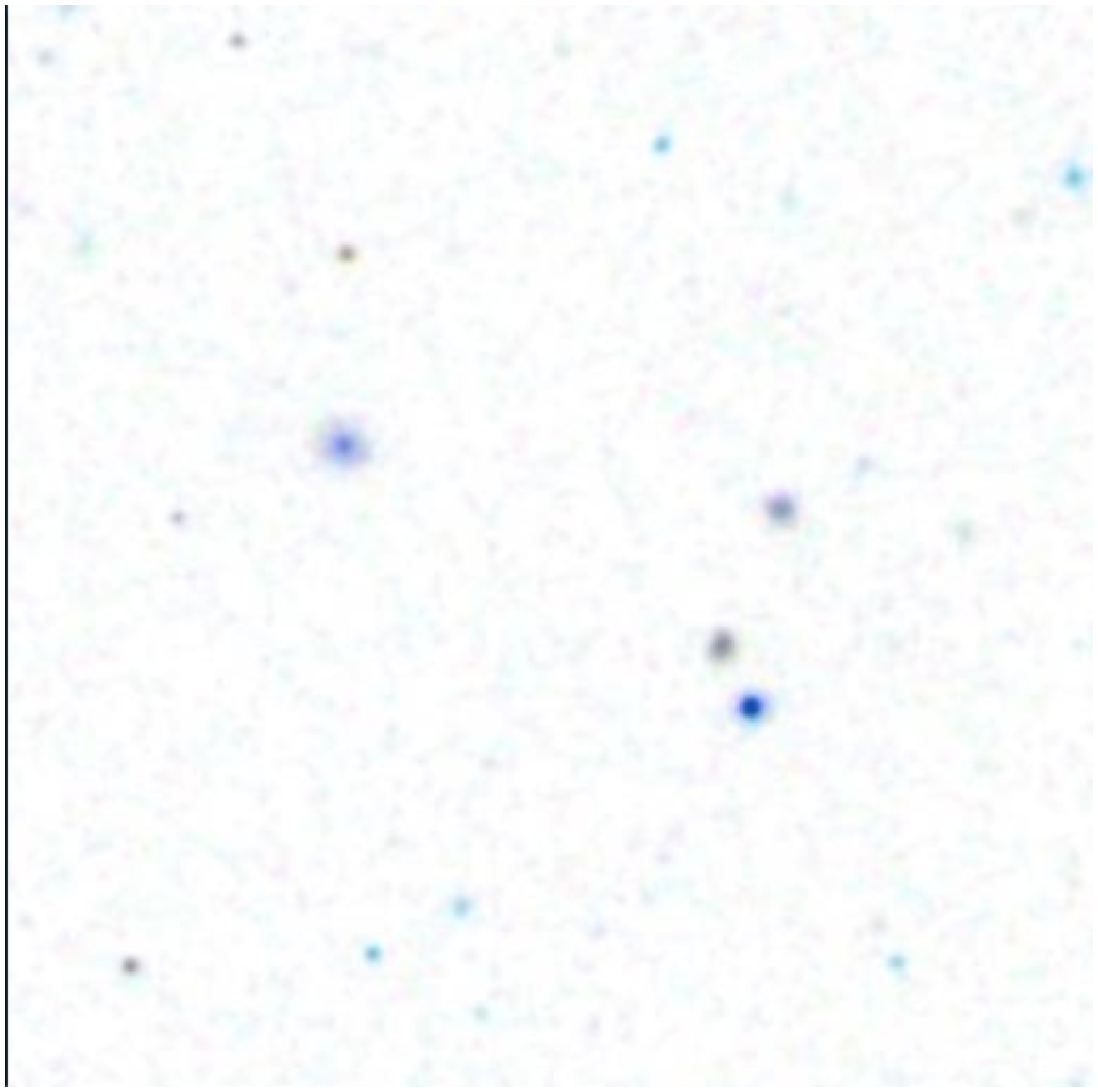}\
 \caption{Inverted SDSS images of the candidate. The left is a larger-field view, and the right image is a zoom-in. No object is seen
at the expected position.}
 \label{SDSSimages}
\end{figure*}

Pictures of the candidate in the POSS-1 Band E (image taken 1950-03-16, resolution 1.6''/pixel), POSS-2 Band F 
(image taken 1992-03-10, resolution 1.0''/pixel) and SDSS and are shown in Figures \ref{POSSimages} and \ref{SDSSimages}. 
They are taken directly from CDS Portal \footnote{http://cdsportal.u-strasbg.fr}. We can see the object clearly
in the POSS-I image. The object is still visible very faintly in the red survey of POSS-II, 
but no longer in the SDSS. Another view is through the interface of The STScI Digitized Sky Survey
\footnote{http://stdatu.stsci.edu/}, included with 2 $\times$ 2 arcmin large zoom-ins of the same
region.

For the candidate, we search for possible counterparts within 5 arcseconds in the infrared surveys 2MASS, 
WISE and AKARI and the VizieR catalogue. An infrared counterpart could indicate that the lost USNO-B object is physically present at the expected position but was undetected in the SDSS. However, if the lost USNO-B object that previously was detected is now only observable in the infrared, that is also a good cause for 
speculation -- could perhaps a Dyson sphere have been built around the star during these 60 $-$ 70 years?
However -- the object cannot be found in any of the other surveys, which excludes this possibility.
Also in the intermediate Palomar Transient Factory public survey \footnote{http://irsa.ipac.caltech.edu/applications/ptf/}
in a sample of images taken from 2009-2010 in the R-band and G-band, nothing is seen.

\begin{table*}[ht]
\caption{The candidate. The table contains information from USNO-B1.0 for the candidate.}
\centering
\begin{tabular}{c c}
\hline\hline
\multicolumn{2}{c}{Properties of the candidate} \\
\hline\hline

USNO-B1.0 name & 1084-0241525\\
ra & 224.4024\\
dec & 18.41725\\
variable? & no \\
m$_{Blue,1st}$ & 21.03\\
m$_{Blue,2nd}$ & \\
m$_{Red,1st}$ & 19.74\\
m$_{Red,2nd}$ & 19.91\\
star/galaxy class 1st Blue & 7\\
star/galaxy class 2nd Blue & -\\
star/galaxy class 1st Red & 7\\
star/galaxy class 2nd Red & 8\\
proper motion, m.a.s./year & 0\\
\hline\hline
\end{tabular}
\label{Candidates}
\end{table*}

Assuming this candidate eventually is found with follow-up photometry, we determine the probability $p$ to detect a 
truly disappearing object from the survey during a decade (the time difference between the POSS-II and SDSS mean epochs) to be less 
than $p < 10^{-6}$. But since impossible effects are not a necessary condition for a super-civilisation
to exist (only an indicator of super-civilisations if finally detected) the probability limit should not be translated
into a limit on the prevalence of super-civilisations. Moreover, the probability is not 
equal to the total probability of observing an object disappear in the Milky Way. Considering 
the age of a galaxy ($\sim$ 10$^{10}$ years), a decade is an extremely short time frame and brings a large 
uncertainty into the given estimate. The total number of stars in the Milky Way (N$_{MW}$ $\sim$ 10$^{11}$) 
is much larger than the number of stars in any of our samples (N $\sim $10$^{6}$). This means a large 
number of stars would need to disappear in the Milky Way during this short time for us to have a chance of noticing 
the effect in our samples that cover just a small fraction (fraction f$\sim$ 10$^{-5}$) of the entire 
set of stars. Therefore, even if a Milky Way star disappeared between the POSS and SDSS surveys, 
it is unlikely we will notice it in our samples. To set a truly realistic limit on observing the 
disappearance of an object in the Milky Way, we need much larger samples.

\subsection{Efficiency of the method}

Small errors in astrometry might give slight differences in coordinates
in USNO-B1.0 and SDSS. Even when SDSS claims to have found a match for the
USNO-B1.0 object, it is not necessarily the same object that is found
at the given position. Its properties are shown in Table \ref{Candidates}.

In order to correctly estimate the upper bound on the probability of detecting 
a missing object-event, we must estimate the efficiency of the successful
matches, for example if only 15\% of all successful matches target the initial
USNO-B1.0 object, it means that the lower bound of the probability 
is highly underestimated by a factor of 7.

We want to examine if it is the same objects we have found in both surveys
at the given positions. We take 10 galaxies with SDSS matches within 1.2 arcseconds (742913 objects) 
from the NDETVar sample and compare the relative positions of the objects in the 
fields in the POSSII/UKSTU Red Survey (via STScI Digitized Sky Survey) and SDSS. 
Indeed, 10 out of 10 objects are the same ones, showing the robustness of the method.

However it should be noted that when looking for variability in galaxy
samples from \cite{VillarroelNyholm2015} through the SDSS Casjobs interface (Table: USNO), 
we find that about 90\% of the galaxies at $z$ $<$ 0.2 show variability on scales $|\Delta r | > $ 0.25 mag and/or
 $|\Delta b | > $ 0.25 mag in the USNO-B1.0 -- an unphysical result. It is possible that 
magnitude measurements for extended objects in the USNO catalogue have big uncertainties due to imprecise targeting.
For point-like objects, it will be less likely.

\section{Discussion}

Using measurements from three different epochs, we have searched for disappearing objects in the sky
during the last decade. We find one candidate object, although an uncertain one due to its faintness
in POSS-2. Improved image analysis is needed to determine with confidence whether it is a real detection,
as was originally reported in USNO-B1.0.

If real, the object could be an extreme, but natural anti-transient: a cataclysmic variable, 
an eclipsing binary \citep{Lipunov} or a highly variable quasar whose luminosity fell below the detection 
limit of the SDSS. If so, follow-up studies with more powerful telescopes surely will reveal its presence 
at the expected position and teach us more about the nature and physics of the object. But if it is not found even with the largest telescopes, this opens up for fascinating, new interpretations.

If the object eventually is reobserved, it sets the probability of identifying entirely vanishing objects to 
be less than one in one million during the last decade with the given samples. This is not so little -- our own Galaxy harbours 
about 400 billion stars, USNO-B1.0 contains information for 1 billion objects and the upcoming Gaia will bring new possibilities with
its improved astrometry for Milky Way stars. Now we have only studied objects with very small proper motions, 
neglecting stars close to the Solar System. But if we use all objects in the USNO-B1.0 catalogue with at least four detections (about 40\% of the entire catalogue) and with accurately measured coordinates and proper motions, we can use a similar method to visually sort out objects that genuinely are missing in images from the Sloan Digital Sky Survey, APASS, Pan-Starrs and iPTF surveys. The positions of the preliminary targets can be further cross-matched against the first Gaia data release. With the multitude of surveys taken during slightly different years and the large number of images one must examine, this gives opportunity for an excellent citizen science project where volunteers only have to determine if an object is seen or not at the center of an image. Citizen science has already led to fascinating discoveries e.g. the discovery of the unusual light curve of 
KIC 8462852 \citep{Boyajian}. Finally, we hope for a fresh remake of the Sloan Digital Sky Survey in the future, adding one epoch of observations, so that other hypothetical ``impossible'' effects can be probed and examined by comparing the properties of objects in the current SDSS and the reobserved SDSS. The full potential of this study waits to be explored with the Large Synoptic Survey Telescope and improved methods of identifying lost objects from our ever-changing sky.



\section{Acknowledgments}

B.V. wishes to thank first and foremost E. Zackrisson, E. Stempels, A. Nyholm, 
and S. Lepine for helpful advice during the project. She also wishes
to thank A. Korn, E. Aronsson, Pianist, A. Gupta, A. Magnard, P. de Sarasate and 
R. Oppenheimer for inspiring discussions. A big thanks also goes to J. Silvester
and A. Gavel for proof-reading the manuscript and, finally, the anonymous referee for constructive 
and helpful comments.

Funding for the creation and distribution of the SDSS Archive  has  been  provided  by  the  Alfred  P.  Sloan
Foundation, the Participating Institutions, the National Aeronautics  and  Space  Administration, the  National
Science Foundation, the US Department of Energy, the Japanese  Monbukagakusho, and the Max-Planck-
Gesellschaft. The SDSS Web site is http://www.sdss.org/. The SDSS is managed by the Astrophysical Research Consortium for the Participating Institutions. The Participating Institutions are the University of Chicago, Fermilab, the
Institute for Advanced Study, the Japan Participation Group, Johns Hopkins University, Los Alamos National
Laboratory, the Max-Planck-Institut f\"ur Astronomie, the
Max-Planck-Institut f\"ur Astrophysik, New Mexico State University, the University of  Pittsburgh, Princeton
University, the US Naval Observatory, and the University of Washington.



\end{document}